\documentclass[iop,apj]{emulateapj}

\newcommand       \mum          {\,{\rm \mu m}}

\begin{document}
   \title{Dimming and CO absorption toward the AA Tau protoplanetary disk: An infalling flow caused by disk instability?} 

\author{Ke Zhang\altaffilmark{1},  Nathan Crockett\altaffilmark{2},  Colette Salyk\altaffilmark{3}, Klaus Pontoppidan\altaffilmark{4}, \\
Neal J. Turner\altaffilmark{5}, John M. Carpenter\altaffilmark{1}, Geoffrey A. Blake\altaffilmark{2}
}

\altaffiltext{1}{Division of Physics, Mathematics \& Astronomy, MC 249-17, California Institute of Technology, Pasadena, CA 91125, USA; kzhang@astro.caltech.edu}
\altaffiltext{2}{Division of Geological \& Planetary Sciences, MC 150-21, California Institute of Technology, Pasadena, CA 91125, USA}
\altaffiltext{3}{National Optical Astronomy Observatory, 950 North Cherry Avenue, Tucson, AZ 85719, USA}
\altaffiltext{4}{Space Telescope Science Institute, 3700 San Martin Drive, Baltimore, MD 21218, USA}
\altaffiltext{5}{Jet Propulsion Laboratory, California Institute of Technology, Pasadena, CA 91109, USA}

  \begin{abstract}

AA Tau, a classical T Tauri star in the Taurus cloud, has been the subject of intensive photometric monitoring for more than two decades due to its quasi-cyclic variation in optical brightness. Beginning in 2011, AA Tau showed another peculiar variation -- its median optical though near-IR flux dimmed significantly, a drop consistent with a 4-mag increase in visual extinction. It has stayed in the faint state since.
 Here we present 4.7\,$\mum$ CO rovibrational spectra of AA Tau over eight epochs, covering an eleven-year time span, that reveal enhanced $^{12}$CO and $^{13}$CO absorption features in the $J_{\rm low}\leqslant$\,13 transitions after the dimming.  These newly appeared absorptions require molecular gas along the line of sight with T$\sim$500\,K and a column density of log (N$_{\rm ^{12}CO}$) $\sim$18.5\,cm$^{-2}$, with line centers that show a constant 6\,km s$^{-1}$ redshift. The properties of the molecular gas confirm an origin in the circumstellar material. We suggest that the dimming and absorption are caused by gas and dust lifted to large heights by a magnetic buoyancy instability.  This material is now propagating inward, and on reaching the star within a few years will be observed as an accretion outburst.  \end{abstract}

   \keywords{
                protoplanetary disks --stars: individual (AA Tau)--stars: pre-main sequence -- stars: variables: T Tauri, Herbig Ae/Be
               }

   \maketitle


\newpage

\section{Introduction}
\label{sec:intro}

Sun-like stars are born surrounded by disks which regulate the angular momentum redistribution required for accretion. It is also in these disks that planets form. Understanding the angular momentum transport and accretion processes in  circumstellar disks is thus critical for our understanding of star and planet formation \citep{Turner14}. Photometric and spectroscopic monitoring of young stars and their accretion disks is an important way to study the dynamics at play. 

The classical T Tauri star AA Tau has provided many clues to the magnetospheric accretion process that channels material from the inner edge of the disk onto the central star \citep{Bouvier99}. AA Tau shows a quasi-cyclic variation with a period of 8.5 days in its optical photometry with an amplitude of 1.4
mag in the $BVRI$ bands \citep{Bouvier99, Bouvier03, Bouvier07, Menard03, Grankin07}.  This peculiar behavior is interpreted as a result of periodic occultations of the star by a non-axisymmetric inner disk warp, perhaps driven by a misalignment between the stellar magnetic dipole and rotation axes \citep{Bouvier99, Bouvier03, Terquem00, OSullivan05, Donati10, Esau14}.  Such occultations require that the AA Tau disk has a nearly edge-on ($i\sim71-75$\arcdeg) orientation, which has indeed been seen in optical scattered light images \citep{Cox13}. 

\begin{figure}[htbp]
\begin{center}
\includegraphics[width=3.5in]{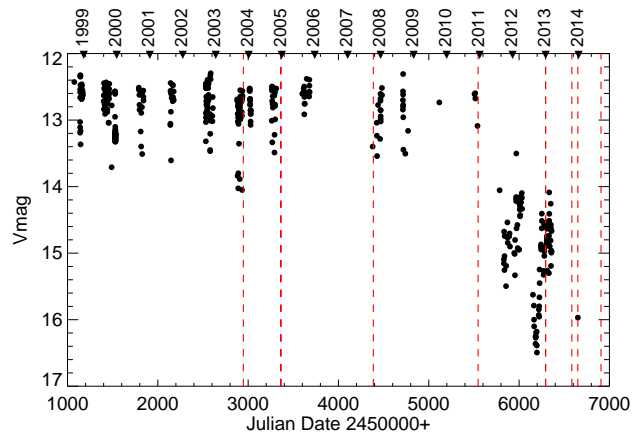}
\caption{$V$-band photometry time series of AA Tau \citep{Bouvier13} overlapped with observational dates of CO $M$ band spectra (red vertical dash lines). The 2013 December photometry data are from Grankin and Bouvier (priv. commun.). }
\label{fig:vband}
\end{center}
\end{figure}

Recently, another peculiar variation was discovered \citep{Bouvier13}. In 2011, AA Tau dimmed  $\sim$ 2 - 4 mag in the $V$ band after maintaining a $V$-band magnitude between 12.3 and 14.3 mag for more than two decades (1987-2010, \citealt{Grankin07}, see Figure~\ref{fig:vband}). Its visual extinction increased by 4 mag based on the $JHK$ color changes, and AA Tau has remained in the faint state since (September 2014, current $V$-band mag=14-17, \citealt{Bouvier13}, and Grankin, priv. commun.).  Surprisingly,  no significant change was found in the mass accretion rate onto the central star by comparing accretion tracers observed before and one year after the dimming started (based on H$\alpha$, H$\beta$ and [HeI] line fluxes, \citealp{Bouvier13}). This is in contrast to other  observations of disk variability, which are usually accompanied by accretion rate changes \citep{Sitko12}.  \citet{Bouvier13} concluded that the sudden dimming is due to additional circumstellar extinction along  the line of sight and suggested that the extra extinction is possibly produced by a non-axisymmetric overdense region in the outer disk  which recently moved into the line of sight via disk Keplerian rotation.  
 Despite the importance of discovering the origin of the enhanced extinction, photometric data alone provide limited insight into the physical and dynamical properties of the extra material along the line of sight, so we turn to spectroscopy to measure gas kinematics and temperature.

Here we present new observational constraints on the origin of the dimming of AA Tau in the form of eight epochs of CO 4.7\,$\mum$ spectra taken over an eleven-year time span. Significantly enhanced $^{12}$CO and  $^{13}$CO absorption components are seen in the spectra taken after the dimming started, at a velocity redshifted with respect to the star. The spectra can be used to measure the properties of the absorbing gas, providing new constraints to the origin of both the dimming and absorption. Indeed, since the optical extinction and CO absorption appear at similar times, and have persisted since, the absorbing gas and obscuring dust most likely arise from the same event(s). We argue that a non-axisymmetric region in the Keplerian disk is unlikely to account for the extra absorption and velocity shift, and suggest that the enhanced extinction and extra absorption may be produced by a disk instability-driven infalling flow. 

\section{Observations}
\label{sec:obs}

We carried out high resolution spectroscopic observations  of AA Tau at eight different epochs between November 2003 and September 2014.  Figure~1 shows the dates of the $M$-band observations,  overlaid on the $V$-band photometry time series of AA Tau over the last decade. A detailed observation log is provided in Table~\ref{table:log}. Most of our spectra were taken with NIRSPEC \citep{McLean98},  a high resolution spectrometer (R$\sim$25,000) at the Keck II telescope, as part of a large NIRSPEC survey of protoplanetary disks \citep{Blake04, Salyk09}, in both the native seeing and adaptive optics (AO) modes. The $M$-band (4.7\,$\mum$) echelle spectra were observed with a $0.''43\times$24\arcsec\,slit (native seeing observations) or a $0.''041\times2.''26$ slit (with AO).  Two spectral settings were used, covering wavelengths between 4.65 and 5.15\,$\mum$, with gaps between orders. The total wavelength range covers a large portion of  the $v$ = (1-0)  fundamental rovibrational band of $^{12}$CO, i.e., the first two R-branch lines (the total angular momentum quantum number of the lower energy state $J_{\rm low}$=0,1) and the low/mid P-branch ($J_{\rm low}$=1-12 and $J_{\rm low}$=30-40), as well as an accretion tracer, the H\textsc{i} Pf$\beta$ transition. The AA Tau $M$-band spectrum of October 2007 is obtained from archival VLT CRIRES data, taken as part of a large survey on protostars and protoplanetary disks \citep{Pontoppidan11b, Brown13}.  The CRIRES spectrum covers a wavelength range between 4.65 and 4.77$\mum$ at a spectral resolution of $\sim$95,000. 

Objects were observed in nod pairs, with subsequent pairs subtracted from one another to remove telluric emission features. The differenced images were flat field corrected and then averaged to increase the signal-to-noise ratio. We extracted 1-D spectra from the 2-D averaged images using the optimal extraction \citep{Horne86}. The wavelength calibration was derived by fitting a fourth order polynomial to the telluric emission lines within the same echelle order. Telluric absorption features were removed by dividing the AA~Tau spectra by that of HR~1620\,(A7 spectral type), taken close in time and airmass. To investigate the continuum flux variation with time,  we compared the raw photon counts of AA~Tau with those of HR~1620 in the continuum around 4.7\,$\mum$ at different epochs. We found the flux ratio varied less than 30\% from the midpoint of all values obtained over the past eleven years, with no apparent correlation between the continuum flux and CO emission line intensities. 
We thus assumed the 4.7\,$\mum$ continuum is constant and  flux calibrated the 1-D spectra by normalizing the continuum  to the interpolated IRAC fluxes between 3.6\,$\mum$\ (352 mJy) and 4.5\,$\mum$\ (332 mJy)\citep{Luhman06}.  
It is worth noting that we do not expect to detect continuum variability caused by the enhanced extinction, because an A$_V$ of 2-4 mag produces a  flux change of less than 10\% at 4.7\,$\mum$ using a Mathis extinction law \citep{Mathis90}.

\begin{deluxetable}{ccccc}
\tabletypesize{\footnotesize}
\tablecolumns{5}
\tablewidth{0pt}
\tablecaption{AA Tau Observation Log \label{table:log}}
\tablehead{
 \colhead{Obs. Date} &  \colhead{Spectral Range} & Int. time& \colhead{Instrument}& AO$^a$\\
 \colhead{(UT)} &\colhead{($\mum$)}& (min)& &}
\startdata
2003-11-03 & 4.65-4.72, 4.95-5.03 &12&NIRSPEC&N\\
2004-12-27 & 4.65-4.72, 4.95-5.03 &12&NIRSPEC&N\\
2004-12-30 & 4.70-4.78, 5.02-5.09 & 8& NIRSPEC&N\\
2007-10-12 & 4.65-4.77        &24$^b$             & CRIRES &Y\\
2010-12-15 & 4.70-4.78, 5.02-5.09 & 64&NIRSPEC&Y\\
2012-12-31 & 4.65-4.72, 4.95-5.03 &32&NIRSPEC&Y\\
2013-01-01&  4.70-4.78, 5.02-5.09 & 24&NIRSPEC&Y\\
2013-10-19 &  4.65-4.72, 4.95-5.03 &24&NIRSPEC&Y\\
2013-12-23 &  4.65-4.72, 4.95-5.03 &24&NIRSPEC&Y\\
2014-09-06 & 4.65-4.72, 4.95-5.03 & 47&NIRSPEC&N
\enddata
\tablecomments{a: Observations with adaptive optics; b: The final spectrum combines observations with two partially overlapping spectral settings with 10 and 14 min. of integration time, respectively.  }
\end{deluxetable}

\section{Results}
\label{sec:results}
\begin{figure*}
\begin{center}
\vspace{-0.5cm}
\includegraphics[width=7in]{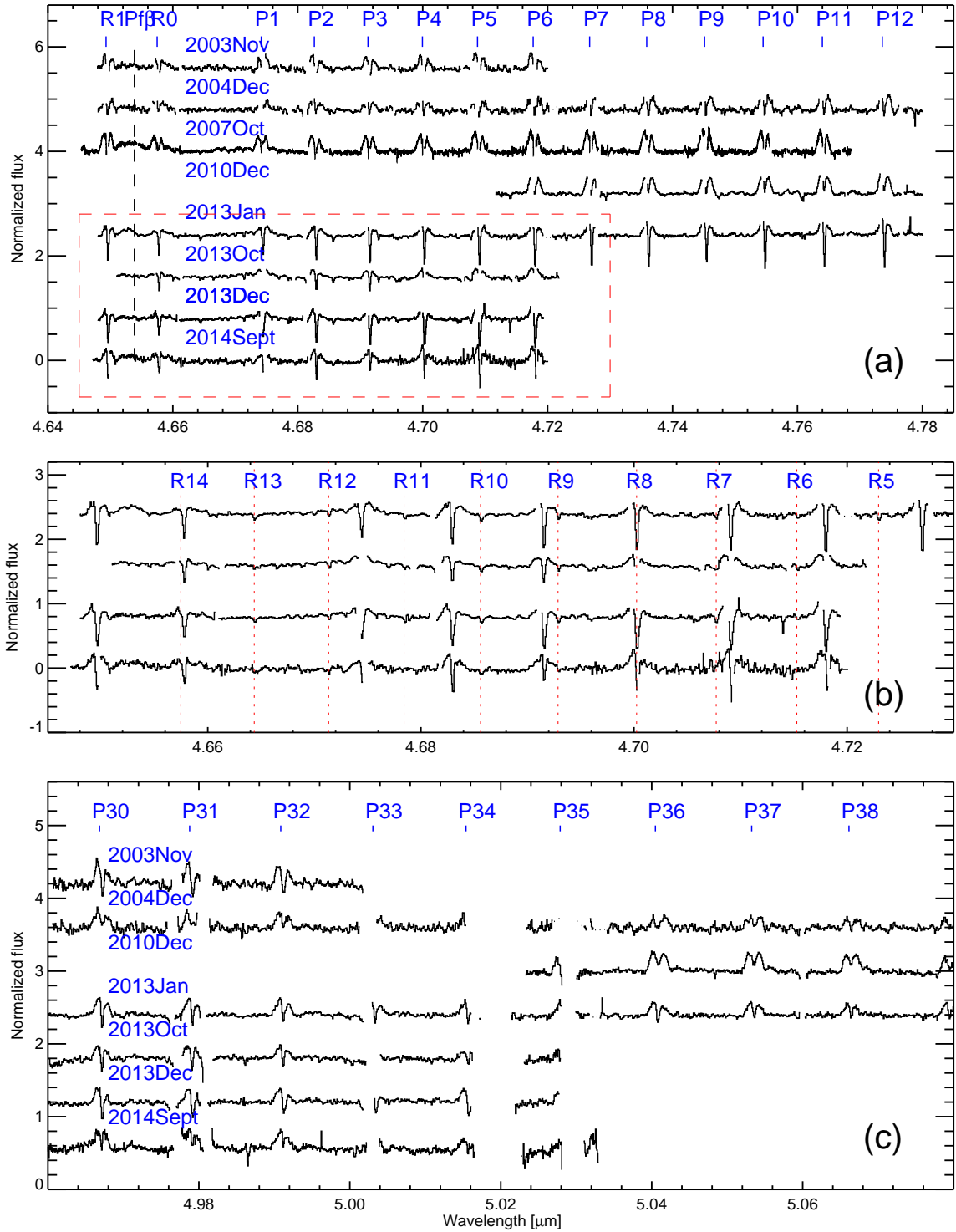}
\caption{CO $M$-band spectra of AA Tau before and after the 2011 $V$-band dimming. (a) CO rovibrational spectra between 4.64 and 4.78\,$\mum$.  The blank regions represent regions with telluric transmission $<$25\%, which we have masked out due to high noise levels. The rest wavelengths of the $^{12}$CO  transitions are marked as small vertical lines at the top of the panel. The region highlighted by the dashed box is enlarged in panel (b). (b) A close up of the 4.64-4.73\,$\mum$ region, highlighting the $^{13}$CO absorption lines that appear after 2011.  Rest wavelengths of the $^{13}$CO $v$= 1-0 lines are marked with dashed lines.  (c) High-$J$ CO spectra between 4.95 and 5.09\,$\mum$.  }
\label{fig:co_spectra}
\end{center}
\end{figure*}

\begin{figure}
\begin{center}
\includegraphics[width=3.5in]{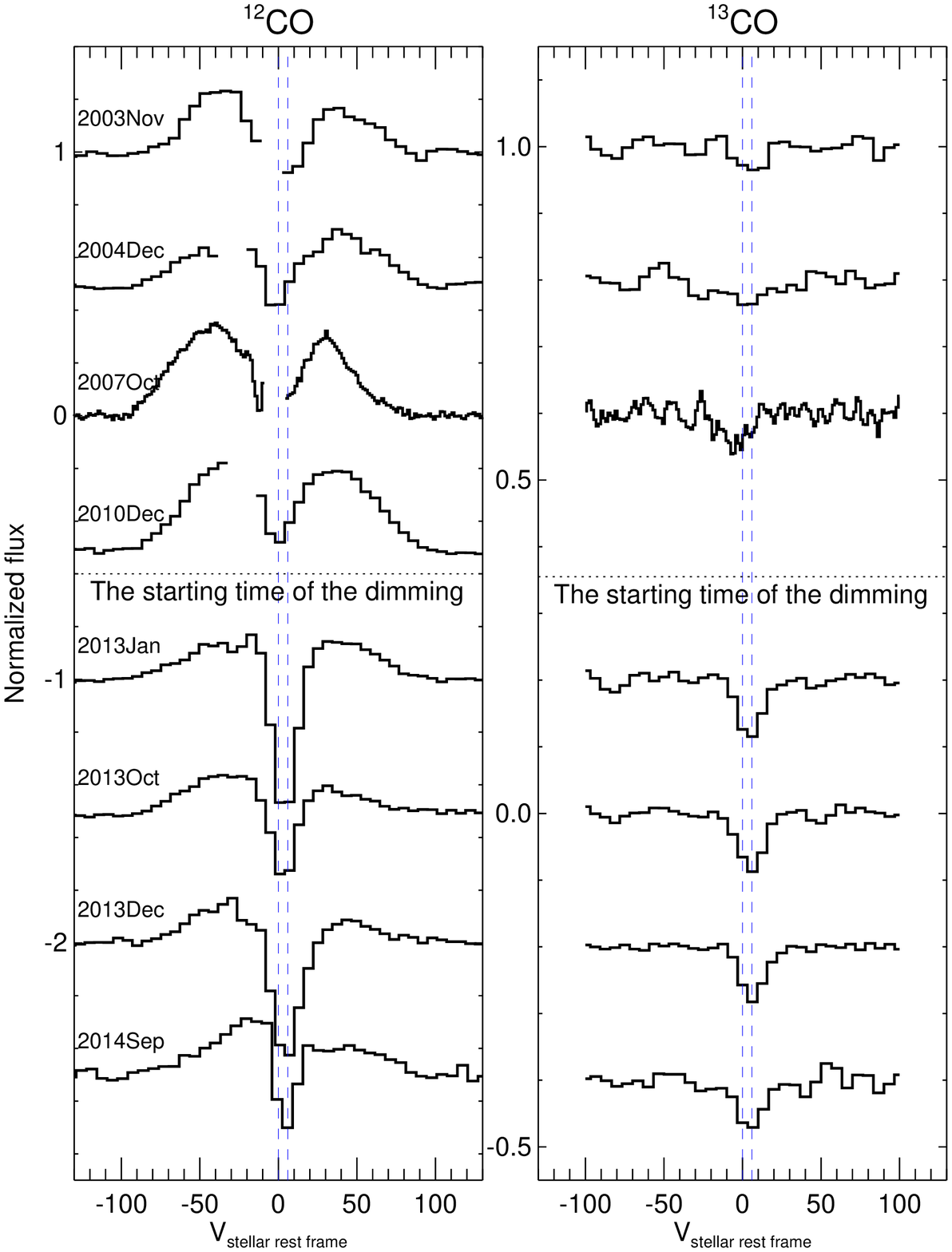}
\caption{ Time variation in the low-$J$ $^{12}$CO and $^{13}$CO line shape(s). The average $^{12}$CO  line shape of each epoch combines the $^{12}$CO $v=$(1,0) $J_{\rm low}\leqslant$12 lines, that for $^{13}$CO averages the isolated from $^{13}$CO R12 and 13 lines. There is no $^{13}$CO line profile for December 2010 epoch because the spectra do not include the two $^{13}$CO transitions.  All of the lines are normalized to the continuum flux. The two vertical dashed lines indicate the locations of stellar velocity and a red shift of 6\,km s$^{-1}$. The horizontal dotted lines denote the onset of dimming in 2011.}
\label{fig:lineshape}
\end{center}
\end{figure}

Figure~\ref{fig:co_spectra} displays our CO $M$-band spectra of AA Tau taken between November 2003 and September  2014 (data obtained within three days were consistent and merged into a single spectrum). Readily detectable $v$ = (1-0) fundamental rovibrational $^{12}$CO emission lines are seen in all eight epochs. The $^{12}$CO emission components appear to be double-peaked and spectrally resolved with an average FWHM of $\sim$130\,km s$^{-1}$. The double-peaked line shape is  expected for emission from a highly inclined, Keplerian disk. By fitting the CO line profile, \citet{Salyk11b} suggested the molecular emission arises from $\sim$0.1\,AU. Our AO data show no CO spectro-astrometric signatures along the disk major axis to a 3$\sigma$ limit of 0.19\,AU, consistent with the previous result.  In five of the eight epochs, the lines were asymmetric, with the blue-shifted peak stronger than the red-shifted peak (Figure~\ref{fig:lineshape}).  The line-to-continuum ratio also varied significantly, from 0.15 to 0.38, but there is no apparent link between the emission line variability and the $V$-band extinction change.

The most striking discovery is the emergence of deep, low- to moderate-$J$ $^{12}$CO ($J_{\rm low}$$\leq$13) and $^{13}$CO  absorption below the continuum in spectra taken after the optical dimming (see Figure~\ref{fig:co_spectra} and \ref{fig:lineshape}). The CO absorption components have a FWHM$\sim$14.2\,km s$^{-1}$, a value close to the width of the nominal  instrumental profile  (FWHM$\sim$12\,km s$^{-1}$).
The mid-$J$ CO lines  between 4.95-5.09\,$\mum$ (of lower state excitation energies  $E_{\rm low}\sim$2000\,K), however, did not show enhanced absorption after the dimming, suggesting the absorbing gas cannot be hotter than $\sim$1000 K. For all the spectra taken before the dimming, the center of the double-peaked emission line profile aligned well with the stellar velocity (heliocentric velocity V$_\star$ = 16.1$\pm$2\,km s$^{-1}$,  \citealt{Hartmann86}). In contrast, the absorption line centers from 2011 and after show a clear and constant redshift of $\sim$6$\pm1$\,km s$^{-1}$. 

Because the goal of this paper is to shed light on the origin of the dimming and absorption towards AA Tau, we focus here on the properties of CO absorption lines. The varying CO emission of AA Tau will be discussed in a separate paper which investigates the variation and asymmetry of $M$-band CO emission in a large sample of disk sources (Crockett et al. 2015, in prep).  We have taken additional L-band NIRSPEC observations before and after the dimming to search for additional molecules such as water in the disk gas; but the complex photospheric structure present at these wavelengths means that only very deep absorption lines would be recognized, and no such features have been found.

 \subsection{The physical properties of the absorbing gas}
\label{sec:abs}

The line profiles of the absorption components are possibly varying (see Figure~\ref{fig:lineshape}), but variations in the Earth's atmosphere can also mimic the line profile change since both the telluric and AA Tau features are marginally resolved, at best. Nevertheless, the line fluxes are less affected and the rotational ladder of CO $v$ = (1-0) lines provides abundant information on the physical properties of the absorbing gas. Under the assumption of a Boltzmann distribution for the rotational states, the line flux ratios provide a direct constraint to the excitation temperature of the gas.  Once the excitation temperature is known, the total column density of the absorbing gas can be derived from the absolute strengths of the absorption components. 

This assumption of local thermodynamic equilibrium is reasonable in this case because the CO $v=0, J\leqslant13$ states are expected to be thermalized due to their moderate critical densities of $\leqslant10^7$ cm$^{-3}$. Given the lack of constraints on the geometry of the absorbing gas,  we make the simplest assumptions, i.e. the gas is isothermal and has a constant number density along the line of sight. Our model thus has three free parameters: 
the gas temperature $T$,  the column density $N$ along the line of sight, and the intrinsic line width $\sigma$ (needed because the $^{12}$CO and $^{13}$CO  absorption lines are largely spectrally unresolved at R=25,000). The line width is an upper limit to the local turbulence velocity since the absorbing gas may exist over a range of radial distances and azimuthal angles. The differential projected radial velocity from gas at different locations can broaden the line profile.

We measured the equivalent width ($W$) of each absorption line as
$W  = \int (I_{\rm cont}-I(v) )/I_{\rm cont}~ dv$.
The  $^{12}$CO equivalent line widths provide a particular challenge since the absorption is built on top of  emission lines. 
Because our goal is the study of the additional absorption that appears after the dimming, 
we measured the equivalent widths of the $^{12}$CO  absorption from the difference between spectra,
specifically those from December 2004 and January 2013  because they cover the largest wavelength range and thus provide the best constraints on the excitation temperature.  Each absorption component in the residual is then fit with a gaussian function rather than directly integrating over the line profile since certain velocities are missing in regions that overlap with saturated atmospheric lines.  The gaussian fits are then divided by the continuum flux to calculate the $^{12}$CO equivalent widths, while those for  $^{13}$CO were measured directly from continuum subtracted January 2013 spectra. 

The theoretical equivalent width of a given transition is calculated using 
$I_v = I_{cont} exp(-sN_J\phi_v)$,
where $N_J$ is the CO column density of a rotational state $J$, $\phi_v$ is the intrinsic line profile, and $s$ is the integrated cross section (cm$^2$ s$^{-1}$), or
\begin{equation}
\label{ }
s = \frac{A_{ul}c^2}{8\pi\nu^2}\frac{g_u}{g_l}[1-e^{-\frac{h\nu}{kT}}]
\end{equation}
The  spontaneous decay rate, $A_{ul}$ is taken from HITRAN (Rothman 2005), while a Voigt function is used for $\phi_v$.  The  Doppler width of Voigt function, $\Delta\nu_D$ = $\sqrt{2}\sigma$, contains $\sigma$ as the third free parameter.  We used an average ISM value of $^{12}$CO/$^{13}$CO number ratio of 69 to scale the column densities of $^{12}$CO to those of $^{13}$CO \citep{Wilson99}.

\begin{figure}
\begin{center}
\includegraphics[width=2.5in, angle=90]{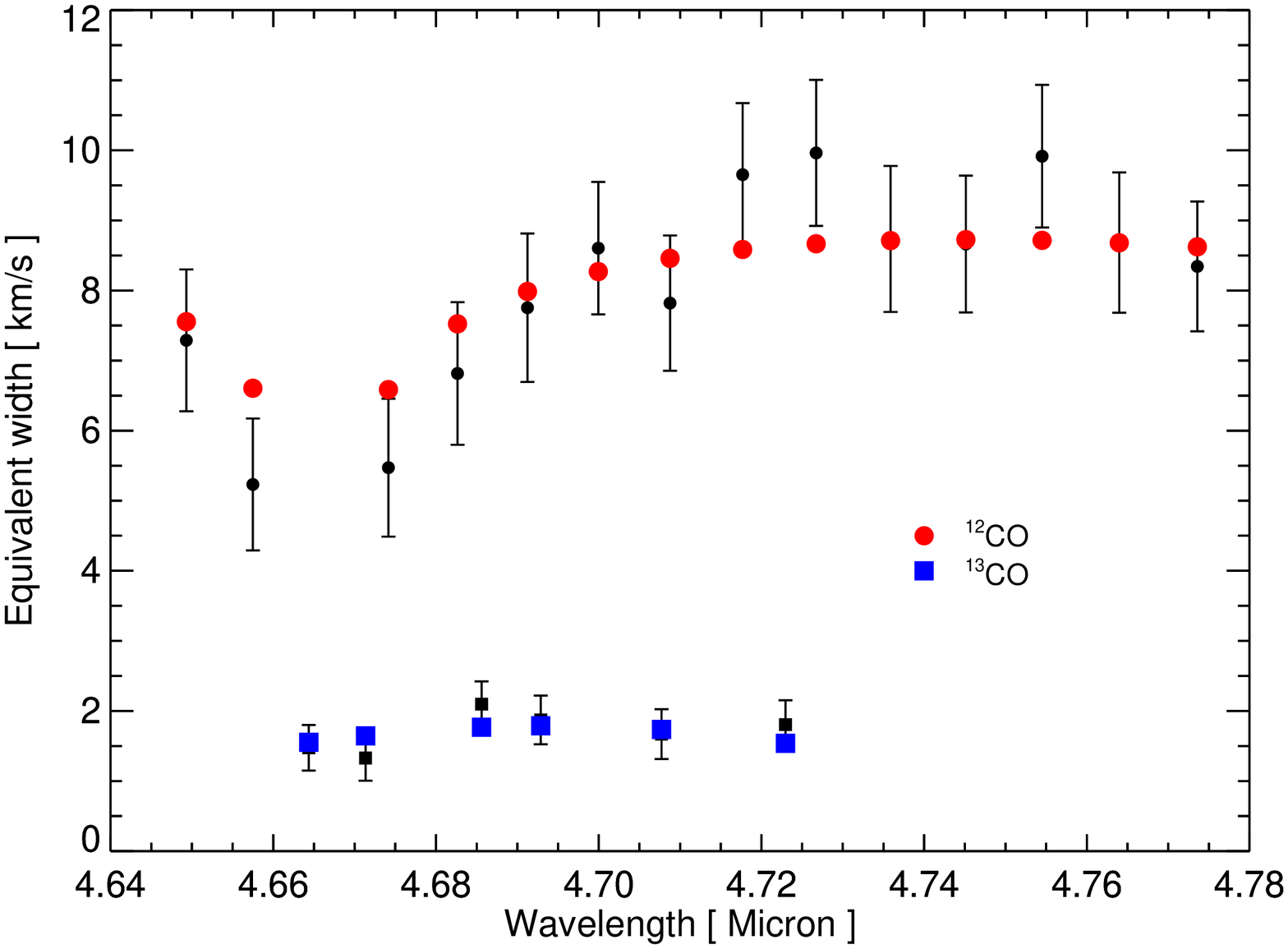}
\includegraphics[width=3.3in]{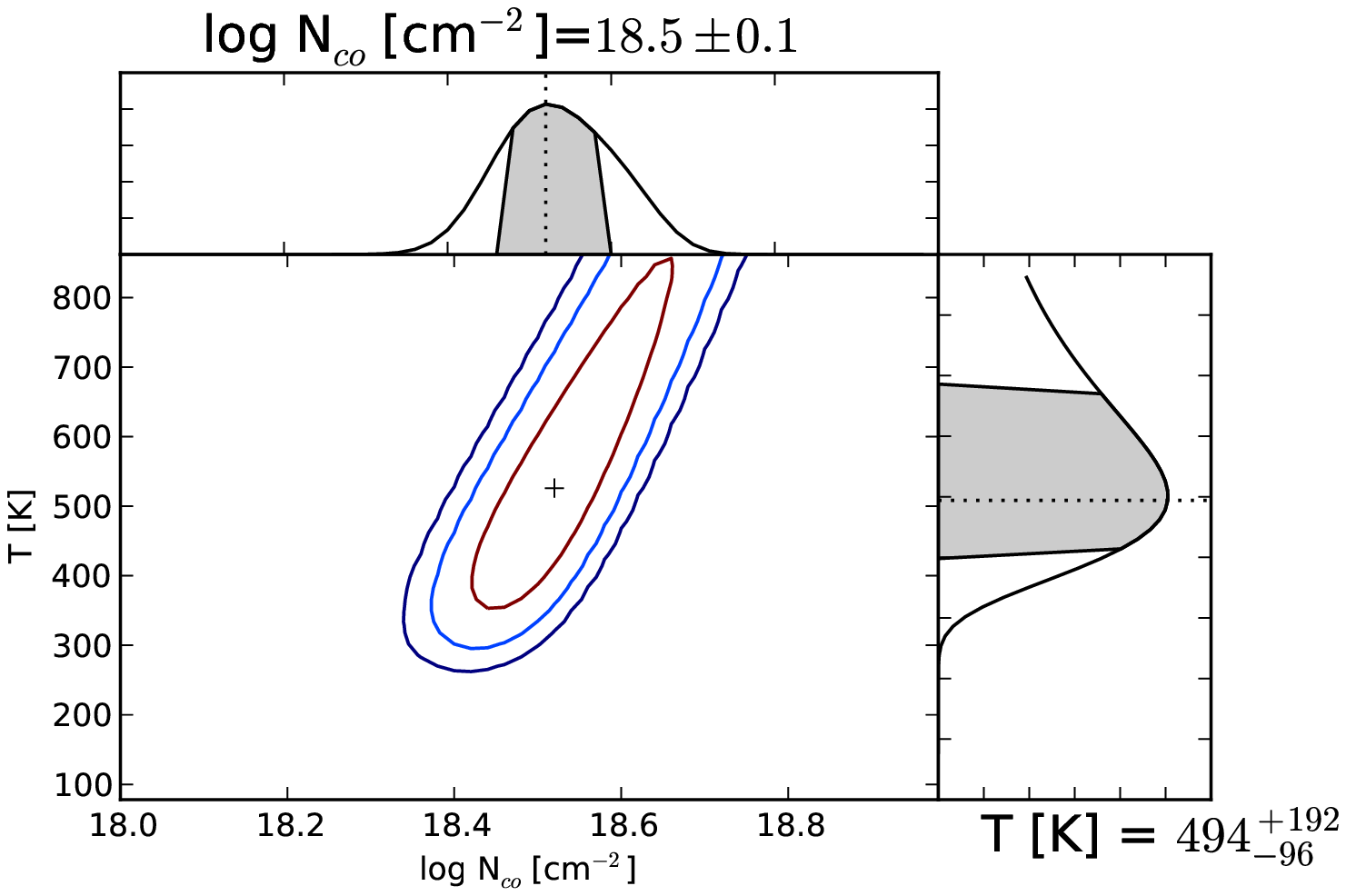}
\caption{\textit{Top:}  Best model of the equivalent widths of the $^{12}$CO and $^{13}$CO absorption components in January 2013. Error bars indicate the 1\,$\sigma$ uncertainties, filled circles $^{12}$CO (black ones are data points and red circles are best model fitting results), filled squares $^{13}$CO (the data points are in black while the model results are in blue). 
\textit{Bottom:} Probability distribution surface contours of T and N$_{^{12} {\rm CO}}$, at 68\%, 95\%, 99\% confidence levels. Marginalized one dimensional posteriors are shown on each side. The shaded regions depict the 68\% confidence areas.}
\label{fig:co_rot}
\end{center}
\end{figure}

The fit results are summarized in Figure~\ref{fig:co_rot}. Our best fit has T = 494$^{+192}_{-96}$\,K, with a total column density of $^{12}$CO along the light of sight of log\,(N$_{\rm ^{12}CO}$) = 18.5$\pm$0.1\,cm$^{-2}$ and an intrinsic width $\sigma$ of 2.2$\pm0.1$\,km\,$\rm s^{-1}$.  We stress that this result is not very sensitive to the input isotopic ratio; changing the $^{12}$CO/ $^{13}$CO value from 69 to 30 would result in T = 427\,K,  log\,(N$_{\rm ^{12}CO}$) = 18.1, and $\sigma$ = 2.6. The best-fit column density decreases in this case because the $^{12}$CO absorption lines are saturated and only the optically thin $^{13}$CO lines provide the main constraints on the column density. A smaller $^{12}$CO/ $^{13}$CO ratio thus produces a smaller N($^{12}$CO).

For the remaining epochs, we did not fit the three parameters ($T$, $N$ and $\sigma$) simultaneously because the available data often cover a smaller wavelength range, with many absorption lines that are heavily contaminated by saturated atmospheric absorption due to unfavorable Earth-induced Doppler shifts.  Since the gas column density is most sensitive to the $^{13}$CO line strengths, we derived the CO gas column density in these epochs by fixing the $T$ and $\sigma$ parameters with the best-fit values from January 2013. In the three 2013 spectra the derived column densities of the absorbing gas vary by $\leq$30\%. For the September 2014 epoch, the column density decreased to log (N${\rm ^{12}CO}$) = 18.2$\pm$0.1\,cm$^{-2}$, about a half of that in 2013. In the pre-dimming epochs, the October 2007 spectrum provides the most stringent column density upper limits, suggesting log (N${\rm ^{12}CO})<17.5$\,cm$^{-2}$ at 99\% confidence ($T$ and $\sigma$ are fixed).  Because the $^{13}$CO lines are optically thin, these column density changes can be directly seen in Figure~\ref{fig:lineshape}.

The contemporary existence of the absorbing gas and the obscuring dust suggests that they are likely to arise from the same event(s). Assuming the gas and dust are co-spatial, we can derive an average gas-to-dust mass ratio in the absorbing material along the line of sight, by combining the $^{12}$CO column density with the increase in visual extinction associated with the dimming.  Assuming a maximum fractional abundance of $n_{\rm CO}/n_{\rm H}$=10$^{-4}$
in protoplanetary disks, the total column density of absorbing gas is at least 3.2$\times$10$^{22}$ cm$^{-2}$, resulting in a $N_{\rm H}/A_V$ $\geq$8$\times10^{21}$ cm$^{-2}$ mag$^{-1}$ -- about a factor of 4 larger than the average ratio measured in nearby molecular clouds (Vuong et al 2003), suggesting a gas-to-dust ratio of $\sim$400 in the absorbing gas. Such an increased gas-to-dust mass ratio is expected if the absorbing gas at large disk heights has experienced significant dust grain growth and/or settling towards the mid-plane. The slope of the spectral energy distribution of AA Tau in (sub)mm wavelength range is consistent with the expectation of significant dust growth \citep{Andrews07}.  

\section{The origin of the absorbing gas}
\label{sec:dis}

The fairly sudden emergence of enhanced extinction and absorption indicates that some obscuring material recently appeared or moved into our line of sight.  We argue that this new material is unlikely to be a foreground source. First, the absorbing gas is too warm ($\sim$ 500\,K) compared to the typical interstellar diffuse cloud excitation temperatures of 10-100\,K \citep{Snow06}. Second, \citet{Bouvier13} found that the enhanced extinction was restricted to AA Tau since three field stars nearby (projected distance between 26\arcsec and 53\arcsec) did not exhibit any luminosity variation. This projected distance is much smaller than that of a typical gas clump in molecular clouds at 140\,pc. For example,  the  \textit{Jeans length} of gas with T of 500\,K and N$_{\rm H_2}$ of 10$^4$ cm$^{-3}$ is 1.3\,pc, i.e.,  a projected distance of 1915\arcsec.

The absorbing gas is also unlikely to be in circumstellar envelope surrounding AA Tau because 
various evolutionary indicators suggest that AA Tau is a Class II object, i.e., a disk with little or no envelope \citep{Robitaille07}. Moreover, gas in a circumstellar envelope should be much cooler than 500\,K \citep{vanDishoec13}.   

The final possibility is that the absorbing material lies in the disk. 
We can place some constraints on its location
based on the gas temperature of $\sim$500\,K. 
Utilizing existing thermo-chemical models in the literature which are constructed for classical T Tauri disks similar to AA Tau,  we can use this temperature to estimate the vertical and radial location of the absorbing gas. In the vertical direction, due to the high inclination of the disk and the optical depth of the infrared CO transitions observed, the absorbing gas must be high in the disk atmosphere but deep enough that sufficient CO column density is present in order to shield itself from photodissociation by high-energy stellar photons \citep{Visser09}.  The dominant heating sources for this warm molecular layer are X-ray and UV radiation (e.g. \citealt{Glassgold04, Gorti11}).  The X-ray luminosity of AA Tau is $\sim10^{30}$\,erg s$^{-1}$ in quiescence and $\sim10^{31}$\,erg s$^{-1}$ during brief flares \citep{Schmitt07, Grosso07}, all within the typical X-ray luminosity of classical T Tauri stars \citep{Feigelson05}. \citet{Najita11} computed a thermo-chemical model for classical T Tauri stars with an $L_X$ of 10$^{30}$\,erg s$^{-1}$. Their model shows that CO starts to survive beyond a vertical column density of N$_{\rm H}\sim10^{21}$\,cm$^{-2}$. At such columns, a temperature of 500\,K requires a radius that is less than 10\,AU distant from the central star. \citet{Walsh10, Walsh12}, for example, used the X-ray, UV continuum and Ly$\alpha$ spectra of TW Hya in their thermal chemical model, and their calculations predict that  the high altitude CO molecular layer inside $\sim$10\,AU is warmer than 500\,K. In summary, the warm temperature of the absorbing gas suggests a location of  $\leqslant$10\,AU in radius.

We conclude with a discussion of the possible origins of the suddenly enhanced extinction and CO absorptions in the AA Tau disk:

(1) \textit{A non-axisymmetric overdense region in the disk}
\vspace{0.05cm}

 \citet{Bouvier13} suggested that the enhanced extinction is caused by a non-axisymmetric structure that recently moved into our line of sight due to Keplerian rotation. However, we argue that this scenario becomes less likely considering the CO absorption components presented here.  Because the $M$-band continuum and the CO emission both arise from very small regions (R$\le$0.5\,AU), only gas and dust located inside a small solid angle can produce the enhanced extinction and absorption. Therefore, the absorption lines should be centered at the stellar velocity. However, the CO absorption lines display a constant redshift of $\sim$6\,km s$^{-1}$ over the past two years. This redshift cannot be explained by the uncertainty of the stellar velocity of AA Tau as all previous measurements of the stellar velocity are consistent within 2\,km s$^{-1}$ (\citealt{Hartmann86} of
16.1$\pm$2\,km s$^{-1}$, \citealt{Bouvier03} of 17.1$\pm$0.9\,km s$^{-1}$ and \citealt{Donati10} of 
17.2$\pm$0.1\,km s$^{-1}$).  A highly eccentric orbit might be able to explain the redshift, but the CO emission lines are well centered at the stellar velocity -- suggesting at least the disk gas in the CO emitting region is in circular motion. If the non-axisymmetric structure rotates at Keplerian speeds, its orbital period is then at least 27 years, the total time span of AA Tau in the bright state and the duration of the dimming. Thus the structure must lie beyond $\geq$8.4\,AU (assuming  M$_\star$=0.8\,$M_\odot$, Bouvier et al. 1999).
 
\vspace{0.1cm}
(2)  \textit{Inward flow driven by disk instability}
\vspace{0.05cm}

Another possibility is the absorbing material has been newly lifted to large heights due to an
    instability of the disk's magnetic fields, leading to an accretion
    outburst involving the inward transport of gas and dust. The buoyant rise of the
   magnetic fields generated in magneto-rotational turbulence has been
   demonstrated in stratified isothermal shearing-box magnetohydrodynmic calculations
   \citep{Miller00}  and in calculations including the stabilizing effect of the external
starlight heating \citep{Hirose11}.  In all cases, the magnetic
fields dominate the pressure above about two density scale heights.
The fields are erratic, with stronger activity coming in bursts, if
the magnetic diffusivity is near the threshold for switching off the
magneto-rotational dynamo (\citealt{Turner07},  \citealt{Simon11a}).

As such material propagates inward, an accretion outburst can potentially result \citep{Zhu10}. Episodic accretion outbursts have been observed in some pre-main sequence low-mass stars (e.g. see a review by Audard et al. 2014). Well-known examples include FU Orionis objects (with typical accretion rates during an outburst  of 10$^{-4}$ M$_\odot$ yr$^{-1}$ and a decay timescale of $\sim$100 years) and EX Lupi objects (smaller but more frequent outbursts with a typical outburst accretion rate of 10$^{-7}$ M$_\odot$ yr$^{-1}$).  Could AA Tau be a pre-FUor/EXor? AA Tau has a quiescent accretion rate between 10$^{-9}$ and 10$^{-8}$ M$_\odot$ yr$^{-1}$, and has shown no significant accretion rate increase since the dimming \citep{Bouvier13}. Similarly, we did not detect significant changes in the accretion tracer H\textsc{i} Pf$\beta$ around 4.65\,$\mum$ before and after the absorption appeared. In Section 3.1, we measured a column density in the extra gas component of 3.2$\times$10$^{22}$ cm$^{-2}$. Assuming this gas is present in a wall at 10\,AU with a scale height of 1\,AU, and that this material will fall onto the star within three years, the accretion rate would rise to $\sim$1.3$\times$10$^{-7}$ M$_\odot$ yr$^{-1}$. Based on the evolutionary stage of AA Tau and our estimate of the accretion rate of the current outburst, we suggest AA Tau is unlikely to be a pre-FUor object but could be a pre-EXor candidate. 

There is one additional hint that suggests episodic outbursts have previously occurred in AA Tau, in that \citet{Cox13} found a chain of Herbig-Haro knots above the AA Tau disk in [SII] images. Such Herbig-Haro objects are usually associated with elevated accretion activity in young stars \citep{Reipurth01}.  The spacing of Herbig-Haro knots in the AA Tau [SII] images, together with
the inferred speed of knots $>$200\,km s$^{-1}$, suggests a time between
ejections of at most a few years, which appears consistent with the smaller and more frequent outbursts in the EX Lupi type.

The observed redshift in the absorption suggests the molecular gas is propagating inward.  The magnitude of the speed is, however, puzzling: 6\,km s$^{-1}$ is too high for a viscous disk but too low for a free-fall velocity.  With a central star of 0.8\,M$_\odot$, the free-fall velocity $\sqrt{2GM_\star/r}$ is 38\,km s$^{-1}$ at 1\,AU. A disk with viscosity $\nu$ will evolve at radius $r$ on a timescale of $t_\nu\sim r^2/\nu$. We adopt the usual parameterization of the viscosity as $\nu=\alpha c_s^2/\Omega$ ($c_s$ is the sound speed, $\Omega$ the Keplerian angular velocity) and use the $\alpha\sim0.1$ found in FU Ori outburst \citep{Zhu07} to estimate a viscous disk timescale.  The radial velocity so obtained is $r/t_\nu\sim \frac{\alpha k T}{\mu m_H} \sqrt{r/GM_\star}$=0.07$\times\frac{\alpha}{0.1}\frac{T}{500\,K}(\frac{r}{AU})^{1/2}(\frac{M_\star}{0.8M_\odot})^{-1/2}$\,km s$^{-1}$.

 The instability driven inward flow scenario is speculative since certain critical disk physical parameters, such as the magnetic field and viscosity, are largely unknown.  Monitoring the physical properties of the absorption components and the accretion rate over the next few years will provide crucial constraints on the origin of the optical dimming.  With a radial velocity of 6\,km s$^{-1}$, matter within 10\,AU from the central star (T$\sim$500\,K) should drive an accretion rate increase over the next few years if the absorption lines are caused by infalling gas. Any subsequent outburst's
accretion rate and duration combined would yield the amount of
material involved, while the duration of the absorption would suggest
a location for the triggering instability.  Furthermore, assuming the
inflowing material rotates at near-Keplerian velocities, the duration
of the absorption constrains the azimuthal extent of the absorber. 
For example, at a distance of 5\,AU, the three-year ongoing absorption suggests an angular extent of at least 90 degrees.  
In short, the dimming and absorption observed in AA Tau may contain important clues into the transport processes in disks.

\begin{acknowledgments}
We thank Jerome Bouvier and Konstantin Grankin for sharing their AA Tau photometric data, and the anonymous referee for helpful comments. KZ, NC and GAB gratefully acknowledge support from the NSF AAG and NASA Origins of Solar Systems programs. CS acknowledges the financial support of the NOAO Leo Goldberg Fellowship program. NJT's contributions were made at the Jet Propulsion Laboratory,
California Institute of Technology, under contract with NASA and with
support from Origins of Solar Systems Program grant 13-OSS13-0114. JMC acknowledges support from NSF award AST-1140063. The spectra presented herein were obtained at the W. M. Keck Observatory, which is operated as a scientific partnership among the California Institute of Technology, the University of California and NASA. The Observatory was made possible by the generous financial support of the W. M. Keck Foundation.  The VLT data presented were acquired under program ID 179.C-0151. Finally, the authors wish to acknowledge the significant cultural role of the summit of Mauna Kea.
\end{acknowledgments}

\bibliographystyle{apj}
\bibliography{ms}

\end{document}